\documentclass[a4paper,twocolumn,fleqn]{article}

\usepackage{ist}
\usepackage{graphicx}
\usepackage{cite,booktabs,multirow,color} %without subfig
\usepackage{subcaption}
\usepackage{comment}
\usepackage{amsmath}
\usepackage{bm}
\usepackage{color}

\def\vector#1{\mbox{\boldmath $#1$}}

\title{Spectral Reflectance Estimation Using Projector with Unknown Spectral Power Distribution}
\author{Hironori~Hidaka, Yusuke~Monno, and Masatoshi~Okutomi\\Tokyo Institute of Technology, Tokyo, Japan}
\date{} % date has an empty field.

\begin{document} 

\maketitle

\thispagestyle{empty} % prevents the first page to be numbered

%%%%%%%%%%%%%%%%%%%%%%%%%%%%%%%%%%
% Abstract
%%%%%%%%%%%%%%%%%%%%%%%%%%%%%%%%%%

\begin{abstract}
A lighting-based multispectral imaging system using an RGB camera and a projector is one of the most practical and low-cost systems to acquire multispectral observations for estimating the scene's spectral reflectance information. However, existing projector-based systems assume that the spectral power distribution~(SPD) of each projector primary is known, which requires additional equipment such as a spectrometer to measure the SPD. In this paper, we present a method for jointly estimating the spectral reflectance and the SPD of each projector primary. In addition to adopting a common spectral reflectance basis model, we model the projector's SPD by a low-dimensional model using basis functions obtained by a newly collected projector's SPD database. Then, the spectral reflectances and the projector's SPDs are alternatively estimated based on the basis models. We experimentally show the performance of our joint estimation using a different number of projected illuminations and investigate the potential of the spectral reflectance estimation using a projector with unknown SPD.
\end{abstract}

%%%%%%%%%%%%%%%%%%%%%%%%%%%%%%%%%%%%
% Overall Document Guidelines: Head
%%%%%%%%%%%%%%%%%%%%%%%%%%%%%%%%%%%%
%
% ====== Introduction ==================================================
%
\section{Introduction}\label{sec:Intro}

Spectral reflectance, which is defined by the reflectance in the wavelength domain, is a fundamental physical property of a scene or an object and provides much richer information than RGB tri-stimulus values captured by a conventional RGB camera or perceived by human eyes. Owing to the detailed information provided by the wavelength-by-wavelength reflectance, many useful applications based on spectral reflectance have been proposed in various fields such as the archive of historical works~\cite{liang2012advances} and the inspection of food quality~\cite{qin2013hyperspectral}.

Multispectral or hyperspectral imaging has been actively studied to estimate scene's spectral reflectance from the images captured using an imaging device acquiring more than three spectral bands. However, existing multispectral imaging systems are typically based on special hardware equipment such as a set of narrow band filters~\cite{gat2000imaging} and a multispectral filer array~\cite{Monno}, which may increase the cost and complexity of the system.

There are also several approaches for estimating the scene's spectral reflectance from the images captured using a standard RGB camera. The representative approach is a lighting-based approach, which observes multispectral measurements by capturing multiple three-band images while temporally changing light spectrum emissions using such as an LED~\cite{Park}, a flash~\cite{Cui} and a projector~\cite{Han2}. Among existing lighting-based systems, a projector-camera setup has demonstrated its effectiveness and a better trade-off regarding accuracy and cost since it utilizes a standard RGB camera and an off-the-shelf projector without any hardware modifications~\cite{Han2}. Recent study has also demonstrated that a projector-camera setup can be used to acquire the spectral 3D model of an object~\cite{li2019pro}. However, existing projector-based spectral reflectance estimation systems assume that the spectral power distribution~(SPD) of each projector primary is known, which requires additional equipment such as a spectrometer to measure the SPD.

In this paper, we present a spectral reflectance estimation method using a standard RGB camera and a projector with unknown SPD toward a more user-friendly system. In our method, we jointly estimate the spectral reflectance and the SPD of each projector primary based on low-dimensional basis models. For the low-dimensional modeling of the spectral reflectance, it is commonly performed to represent the spectral reflectance by a small number of basis functions obtained from a spectral reflectance database~\cite{Parkkinen,Maloney,Monno2}. However, to the best of our knowledge, there is no existing report that models the projector's SPD by basis functions. Therefore, we model the SPD of each projector primary using spectral basis functions obtained from a newly collected projector's SPD database. Based on the low-dimensional basis models, we build a cost optimization framework to altenately estimate the spectral reflectance and the SPD of the projector primary. Through the experiments, we investigate the performance of our joint estimation using a different number of projected illuminations and discuss the potential of the spectral reflectance estimation using a projector with unknown SPD.

% ////////////////////////////////
\begin{figure*}[t!]
    \begin{center}
          \includegraphics[width=\linewidth]{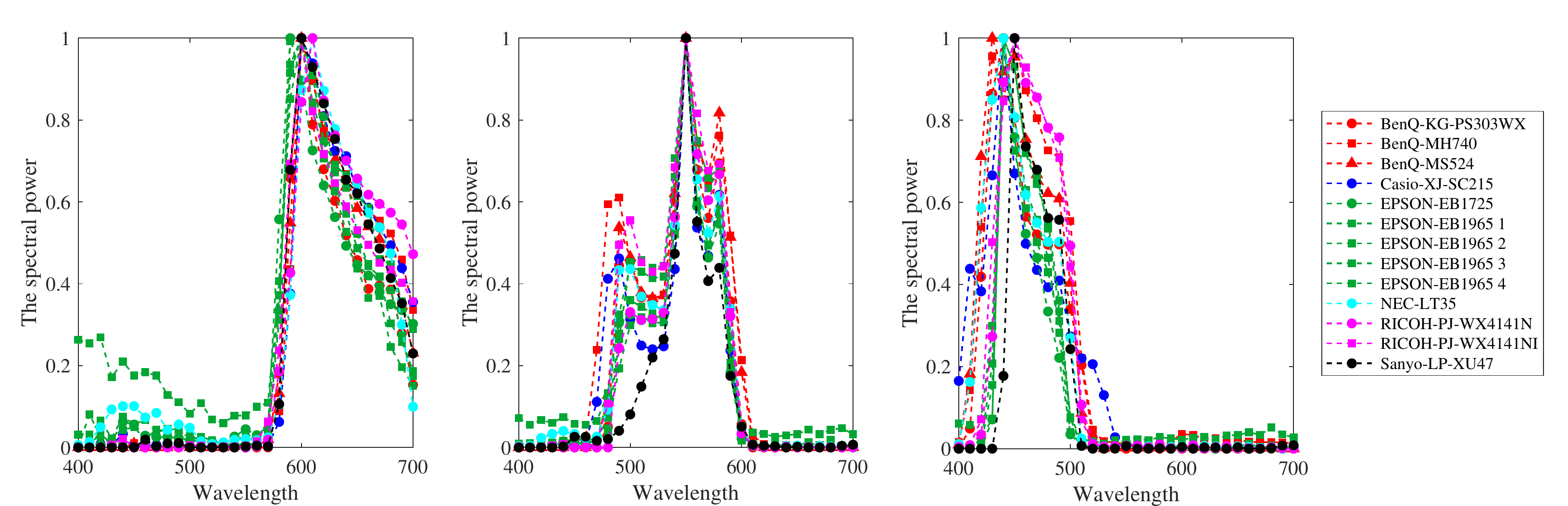}
    \caption{Database of projector's SPDs}
    \vspace{3mm}
    \label{fig:projector_database}
  \end{center}
\end{figure*}

% ////////////////////////////////
\begin{figure*}[t!]
    \begin{center}
    % \begin{tabular}{c}
      \begin{minipage}{0.63\linewidth}
        \begin{center}
          \includegraphics[width=\linewidth]{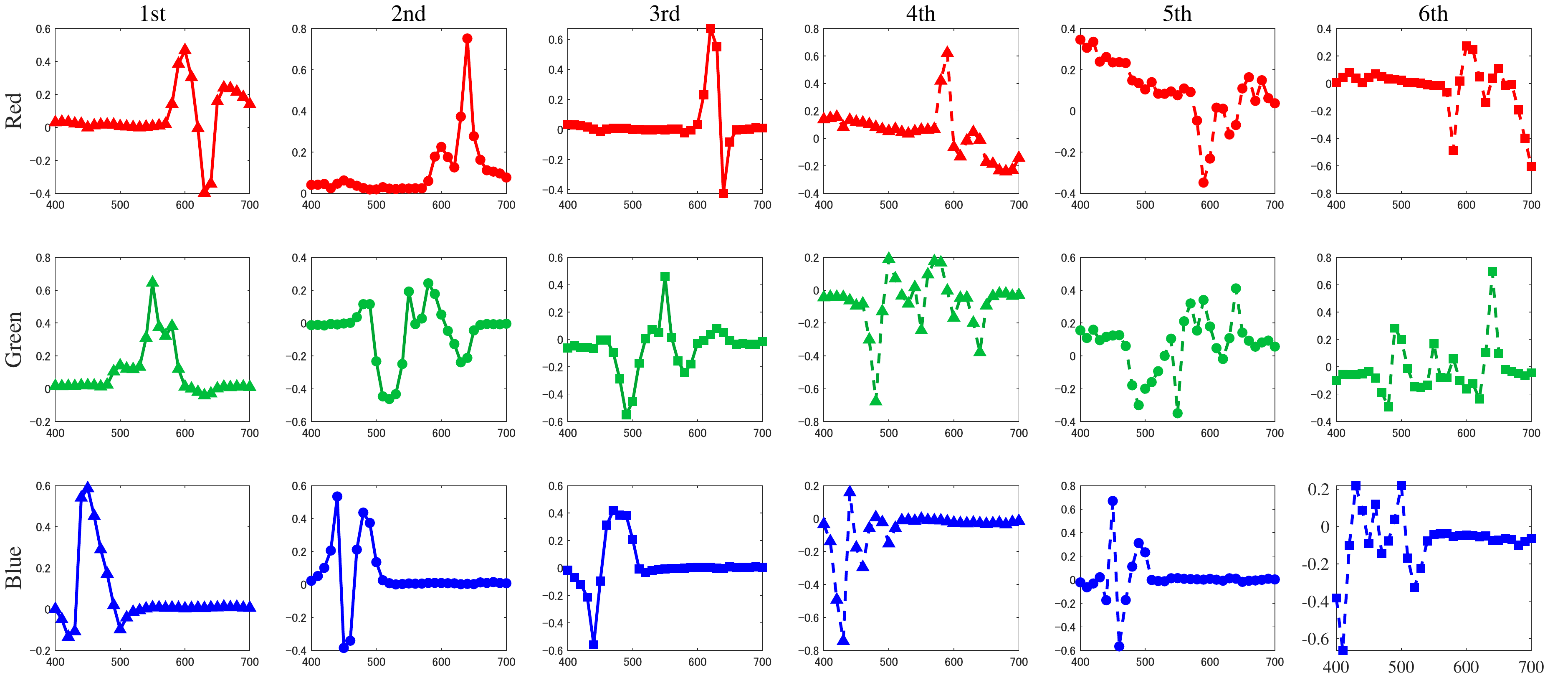}\\ \vspace{-2mm}
        \subcaption{Basis functions for each projector primary}
        \label{fig:basis_ill}
        \end{center}
      \end{minipage}
      \begin{minipage}{0.33\linewidth}
        \begin{center}
          \includegraphics[width = \linewidth]{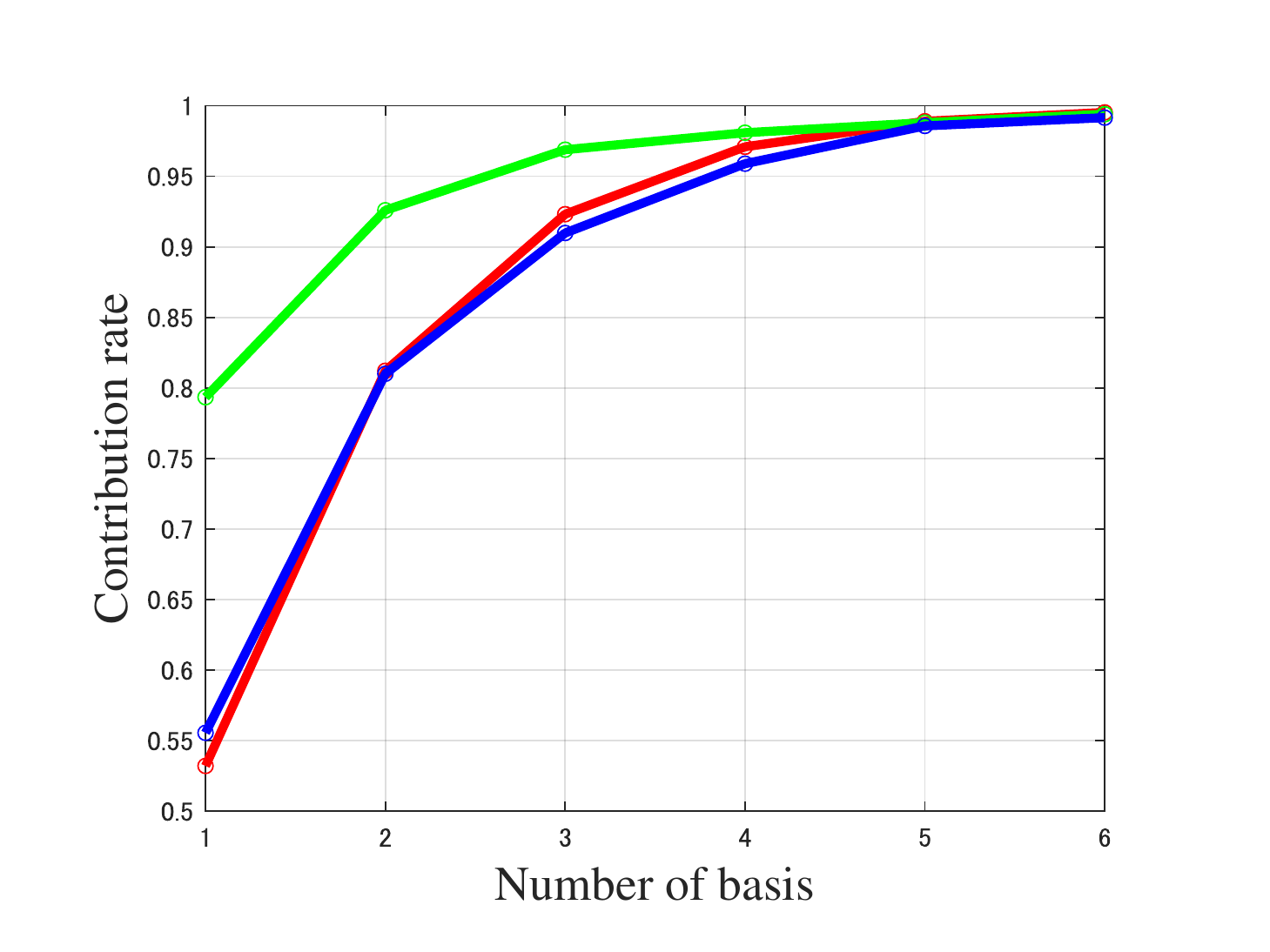}\\ \vspace{-3mm}
        \subcaption{Contribution rate}
        \label{fig:PCArate}
        \end{center}
      \end{minipage}
    % \end{tabular}
    \vspace{8mm}
    \caption{Basis functions of projector illumination}
    \vspace{3mm}
    \label{fig:spectral_basis}
  \end{center}
\end{figure*}

%
% ======= Proposed method ==============================================
%
%%% ============ First part =================
\section{Proposed Method}\label{sec:method}

\vspace{2mm}
\subsection{Image formation model}\label{sec:formulation}

\vspace{1mm}
In our setup, a standard RGB camera and an off-the-shelf projector are used to capture input images. If we assume Lambertian surface reflection and linear camera responses, the pixel value $I$ is represented as
\begin{equation}
    I_{m,n,p} = \int_\Omega c_m(\lambda)s_n(\lambda)r_p(\lambda)d\lambda,
    \label{eq:form}
\end{equation}
where $c_m(\lambda)$ is the camera sensitivity of $m$-th camera channel~($1 \leq m \leq 3$), $s_n(\lambda)$ is the SPD of $n$-th projected illumination by a projector~($1 \leq n \leq N$), $N$ is the number of projected illuminations, $r_p(\lambda)$ is the spectral reflectance of $p$-th pixel position, and $\Omega$ is a target visible wavelength range, which is typically set to [400nm, 700nm]. This equation can be rewritten by a discretized vector form as
\begin{equation}
    \vector I_{m,n,p} = {\bm c}_{m}^{T} diag({\bm s}_n) {\bm r}_p,
    \label{eq:form_matrix}
\end{equation}
where ${\bm c}_{m}$, ${\bm s}_n$, and ${\bm r}_p$ are a camera sensitivity vector, an illumination SPD vector, and a spectral reflectance vector with $N_\lambda$ dimension (e.g., $N_\lambda=31$ if we use the wavelength range of [400nm, 700nm] with 10nm interval discretization), respectively.

It is known that the spectral reflectance of a real-world material is well described by a low-dimensional model using spectral reflectance basis functions as
\begin{equation}
    {\bm r}_p = \sum_{i=1}^{N_r}\alpha_{p,i}{\bm b}_{i}^{ref} = {\bf B}^{ref}{\bm \alpha}_p,
    \label{eq:ref_basis}
\end{equation}
where ${\bm b}_{i}^{ref}$ is $i$-th reflectance basis function, $\alpha_{p,i}$ is the coefficient for $i$-th basis function at $p$-th pixel, $N_r$ is the number of basis functions, ${\bf B}^{ref}$ is an $N_\lambda \times N_r$ basis matrix, and ${\bm \alpha}_p$ is an $N_r$-dimensional coefficient vector for $p$-th pixel. The basis matrix is commonly derived by principal component analysis~(PCA) of publicly available spectral reflectance database, such as the spectral database of Munsell color chips~\cite{Parkkinen,Maloney}.

In this work, we apply a basis model also for projector's illumination to enable the joint estimation of the spectral reflectance and the SPD of each projected illumination based on low-dimensional models. It is known that the SPD of any projected illumination by a projector is typically represented by the sum of the SPDs of three primary illuminations, i.e., red, green, and blue illuminations. Mathematically, this is expressed as
\begin{equation}
    {\bm s}_n = \gamma_{r,n}{\bm s}_{r} + \gamma_{g,n}{\bm s}_{g} + \gamma_{b,n}{\bf s}_{b},
    \label{eq:illumination_RGB}
\end{equation}
where ${\bm s}_{r}$, ${\bm s}_{g}$, and ${\bm s}_{b}$ represent the SPD of red, green, and blue illuminations, respectively, and $\gamma_{r,n}$, $\gamma_{g,n}$, and $\gamma_{b,n}$ represent the gains for each primary to form the SPD of $n$-th projected illumination.

Since there is no public database for the SPD of each projector primary, we collected the SPD of 13 projectors with 10 models using a mercury lamp, as shown in Fig.~\ref{fig:projector_database}. Then, we applied PCA to the collected SPD data to obtain the basis functions for each primary illumination as shown in Fig.~\ref{fig:spectral_basis}(a). From Fig.~\ref{fig:spectral_basis}(b), we can find that the projector SPDs are modeled by six basis functions with more than $99.8\%$ contribution ratios. Using the derived illumination basis functions, each primary illumination is modeled as
\begin{equation}
        {\bm s}_r = {\bf B}_r^{ill}{\bm \beta}_r, \ \ {\bm s}_g = {\bf B}_g^{ill}{\bm \beta}_g, \ \ {\bm s}_b = {\bf B}_b^{ill}{\bm \beta}_b,
        \label{eq:ill_basis}
\end{equation}
where ${\bf B}_r^{ill}$, ${\bf B}_g^{ill}$, and ${\bf B}_b^{ill}$ are $N_\lambda \times N_s$ illumination basis matrices for each primary, $N_s$ is the number of illumination basis functions, and ${\bm \beta}_r$, ${\bm \beta}_g$, and ${\bm \beta}_b$ are $N_s$-dimensional coefficient vectors.

Based on the models of Eq.~(\ref{eq:illumination_RGB}) and Eq.~(\ref{eq:ill_basis}), the SPD of $n$-th projected illumination by a projector is generally modeled as
\begin{equation}
        {\bm s}_n = [{\bf B}_r^{ill},{\bf B}_g^{ill},{\bf B}_b^{ill}]{\bm \Gamma}({\bm \gamma}_n)\left[\begin{array}{c}{\bm \beta}_r\\{\bm \beta}_g\\{\bm \beta}_b\end{array}\right] = {\bf B}^{ill}{\bm \Gamma}({\bm \gamma}_n){\bm \beta}.
\end{equation}
In the above equation, ${\bm \Gamma}({\bm \gamma}_n)$ is a matrix encoding the gains for each primary ${\bm \gamma}_n=[\gamma_{r,n},\gamma_{g,n},\gamma_{b,n}]^T$ and represented as
\begin{equation}
        {\bm \Gamma}({\bm \gamma}_n) = \left(
            \begin{array}{cccc}
                \gamma_{r,n}{\bf I}_{N_s \times N_s} & {\bf O}_{N_s \times N_s} & {\bf O}_{N_s \times N_s} \\
                {\bf O}_{N_s \times N_s} & \gamma_{g,n}{\bf I}_{N_s \times N_s} & {\bf O}_{N_s \times N_s} \\
                {\bf O}_{N_s \times N_s} & {\bf O}_{N_s \times N_s} & \gamma_{b,n}{\bf I}_{N_s \times N_s}
            \end{array}
        \right),
        \label{eq:illumination_RGB1}
\end{equation}
where ${\bf I}_{N_s \times N_s}$ is the ${N_s \times N_s}$ identity matrix and ${\bf O}_{N_s \times N_s}$ is the ${N_s \times N_s}$ zero matrix. By the above form, the SPD of any projected illumination can generally be modeled including the SPDs of three primary illuminations, which are modeled by the cases of $(\gamma_r,\gamma_g,\gamma_b) = (1,0,0), (0,1,0)$, and $(0,0,1)$, respectively.

%%% ============ Second part =================
\subsection{Joint estimation of spectral reflectances and projector’s SPDs}\label{sec:simultaneous}

\vspace{1mm}
Based on the basis models introduced above, we jointly estimate the spectral reflectance and the SPD of each projector primary using $N$ images captured with $N$ projected illuminations. The cost function $E$ is built as 
\begin{equation}
    \begin{split}
        E({\bm \alpha}_p,{\bm \beta})=&\frac{1}{NP} \sum_m\sum_n\sum_p\left( I_{m,n,p}-{\bm \alpha}_p^T {\bf W}_{m,n} {\bm \beta} \right)^2\\
        &+\frac{\sigma_1}{N}\sum_n\lVert \frac{d^2 {\bf B}^{ill}}{d\lambda^2}{\bm \Gamma}({\bm \gamma}_n){\bm \beta} \rVert_2^2+\frac{\sigma_2}{P}\sum_p\lVert\frac{d^2{\bf B}^{ref}}{d\lambda^2}{\bm \alpha}_p \rVert_2^2,
         \label{eq:cost}
    \end{split}
\end{equation}
where $P$ is the number of image pixels and $\lVert\cdot\rVert_2$ represents the L2 norm of the derived vector. The first term is a data fidelity term that evaluates the difference between the observed pixel value and the value obtained by the image formation model with estimated basis coefficients. The second term and the third term are regularization terms that constrain the smoothness of the derived illumination SPDs and spectral reflectances, respectively. The weights for each term are balanced by the parameters $\sigma_1$ and $\sigma_2$. In the data fidelity term, ${\bf W}_{m,n}$ is described as 
\begin{equation}
    {\bf W}_{m,n} = {\bf B}^{ref} diag(\bm{c}_m) {\bf B}^{ill}{\bm \Gamma}({\bm \gamma}_n).
    \label{eq:w}
\end{equation}

To form ${\bf W}_{m,n}$, the gain values $(\gamma_r,\gamma_g,\gamma_b)$ for each projected illumination need to be known. Ideally, these gain values correspond to the input gain values $(\gamma'_r,\gamma'_g,\gamma'_b)$ for each projector primary. However, we experimentally found that this is not always the case, i.e., ${\bm s}_n \neq \gamma'_{r,n}{\bm s}_{r} + \gamma'_{g,n}{\bm s}_{g} + \gamma'_{b,n}{\bf s}_{b}$. Therefore, as in~\cite{Blasinski}, we pre-estimate the gain values based on the observed images with three primary illuminations as
\begin{equation}
        {\bm \gamma}_n = \underset{{\bm \gamma}}{\rm argmin} \left\lVert {\bf I}_{n}-{\bm \gamma}^T
        \left[\begin{array}{c}{\bf I}_r\\{\bf I}_g\\{\bf I}_b\end{array}\right]\right\rVert_F^2
    \label{eq:cost_gamma}
\end{equation} 
where ${\bf I}_{n}$ is the observed image with $n$-th projected illumination, ${\bf I}_{r}$, ${\bf I}_{g}$, and ${\bf I}_{b}$ are the observed images with primary red, green, and blue illuminations, respectively, and $\lVert\cdot\rVert_F$ represents the Frobenius norm of the derived matrix.

Using the matrix ${\bf W}_{m,n}$ with pre-determined gain values ${\bm \gamma}_n$, we minimize the cost of Eq.~(\ref{eq:cost}) to estimate the coefficient vectors, ${\bm \alpha}_p$ and ${\bm \beta}$, subject to the following non-zero and scale constraints.
\begin{equation}
    \begin{split}
        &{\bf B}_r^{ill} {\bm \beta}_r \geq 0, ~~{\bf B}_g^{ill} {\bm \beta}_g \geq 0, ~~{\bf B}_b^{ill} {\bm \beta}_b \geq 0, ~~{\bf B}^{ref} {\bm \alpha}_p \geq 0,\\
       &\left({\bf B}^{ill}{\bm \Gamma}({\bm \gamma}_N){\bm \beta}\right)\left( \lambda_f \right) = 1.
    \end{split}
    \label{eq:constrain_cost}
\end{equation}
In the above constraints, the first four constraints ensure that the resultant illumination SPDs and spectral reflectance do not have negative values, while the last constraint fixes the scale of the estimation by fixing the spectral power of $\lambda_f$ wavelength of $N$-th illumination to one. 

Since the simultaneous optimization of ${\bm \alpha}_p$ and ${\bm \beta}$ is hard to solve, we alternately derive them as performed in~\cite{Oh}.
We start the iteration to first estimate ${\bm \beta}$ (i.e., the illumination SPDs) by setting the first reflectance basis function as the initial spectral reflectance estimates for all the pixels because we experimentally found that the alternating estimation does not converge if we start the iteration to estimate ${\bm \alpha}_p$ (i.e., spectral reflectances).
After ${\bm \beta}$ is estimated to derive the SPD of each primary illuminations, ${\bm \alpha}_p$ is estimated to derive the spectral reflecance of each pixel using fixed ${\bm \beta}$. After that, ${\bm \beta}$ is re-estimated using fixed ${\bm \alpha}_p$. These processes are iterated until the minimized cost converges.

% ////////////////////////////////////////////
\begin{figure}[t!]
    \centering
    \includegraphics[width =0.6\linewidth]{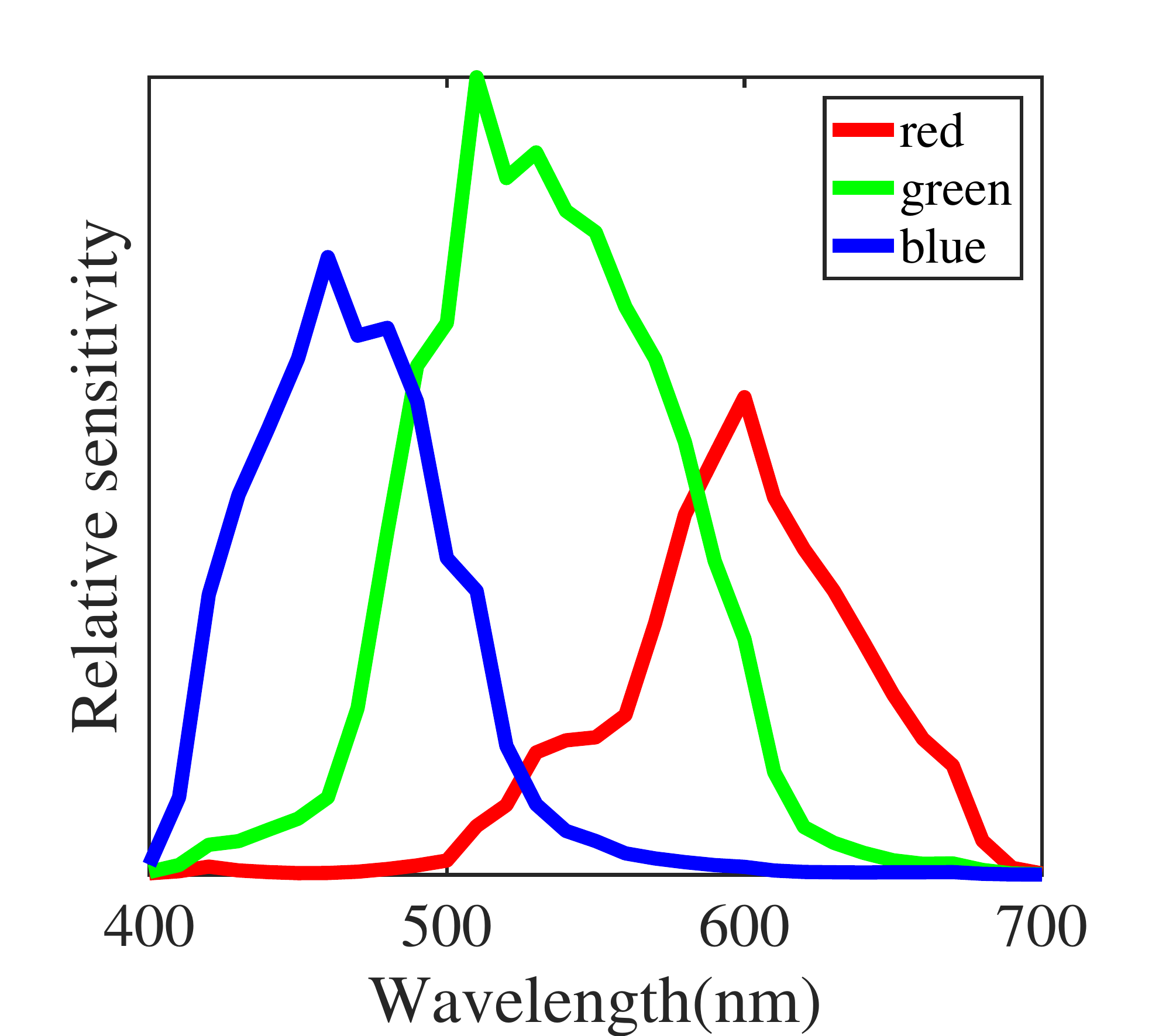}\\ \vspace{1mm}
    \caption{Camera sensitivity functions}
    \vspace{4mm}
    \label{fig:CameraSensitivity}
\end{figure}

%
% ======= Experiments ==================================================
%
\section{Experimental Results}\label{sec:experiments}
\vspace{2mm}
\subsection{Setups}
\vspace{1mm}
In our experiments, we used a Canon EOS 5D Mark II camera and a Casio XJ-SC215 projector. The camera sensitivity of the Canon EOS 5D Mark II model was obtained from the camera sensitivity database of~\cite{Jiang}, as shown in Fig.~\ref{fig:CameraSensitivity}. 

Similar to~\cite{li2019pro}, we captured images under seven color illuminations projected by a projector: red, green, blue, cyan, magenta, yellow, and white illuminations, which were generated by setting input primary gains as $(\gamma'_r,\gamma'_g,\gamma'_b) = (1,0,0)$, $(0,1,0)$, $(0,0,1)$, $(0,1,1)$, $(1,0,1)$, $(1,1,0)$, and $(1,1,1)$, respectively. In the following experiments, we report the performance of joint estimation using different number of projected illuminations.

We used six basis functions for the spectral reflectance, which was obtained by PCA of the specral database of Munsell color chips~\cite{Parkkinen,Maloney}. We also used six basis functions for each projector primary. To estimate the illumination SPDs of the Casio XJ-SC215 projector in our experiment, we removed this projector from the SPD dataset of Fig.~\ref{fig:projector_database} and used recalculated basis functions. The parameters $\sigma_1$ and $\sigma_2$ were empirically set to 0.125 and 0.005, respectively.

% ////////////////////////////////////////////
\begin{figure}[t!]
    \centering
    \includegraphics[width =0.7\linewidth]{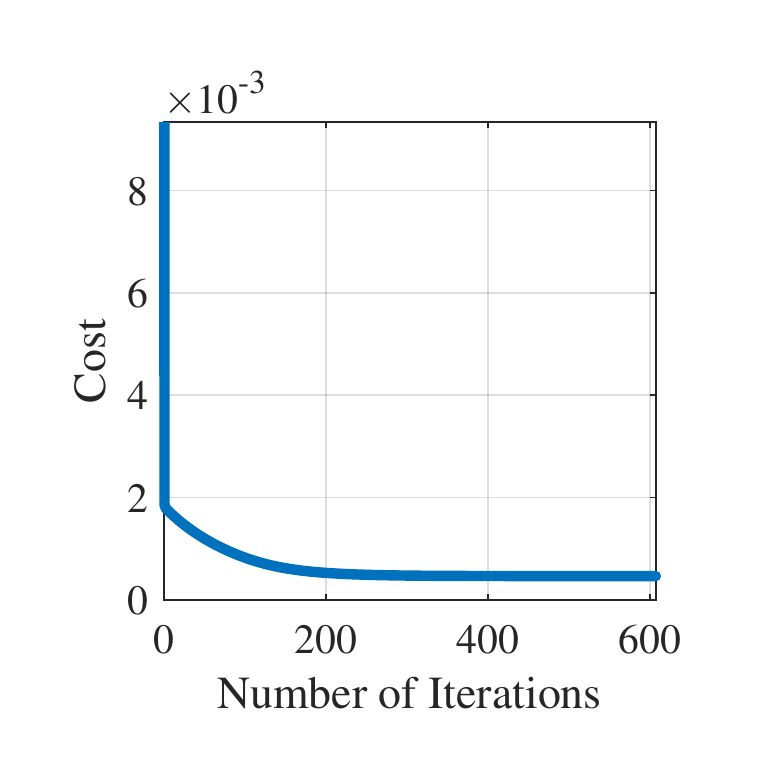}\\
    \vspace{2mm}
    \caption{Cost convergence}
    \vspace{4mm}
    \label{fig:RMSE_variation}
\end{figure}

\begin{figure*}[t!]
\centering
%input images
\begin{minipage}{\hsize}
    \begin{tabular}{c}
    \begin{minipage}[t]{0.133\linewidth}
        \centering
        \includegraphics[width =\linewidth]{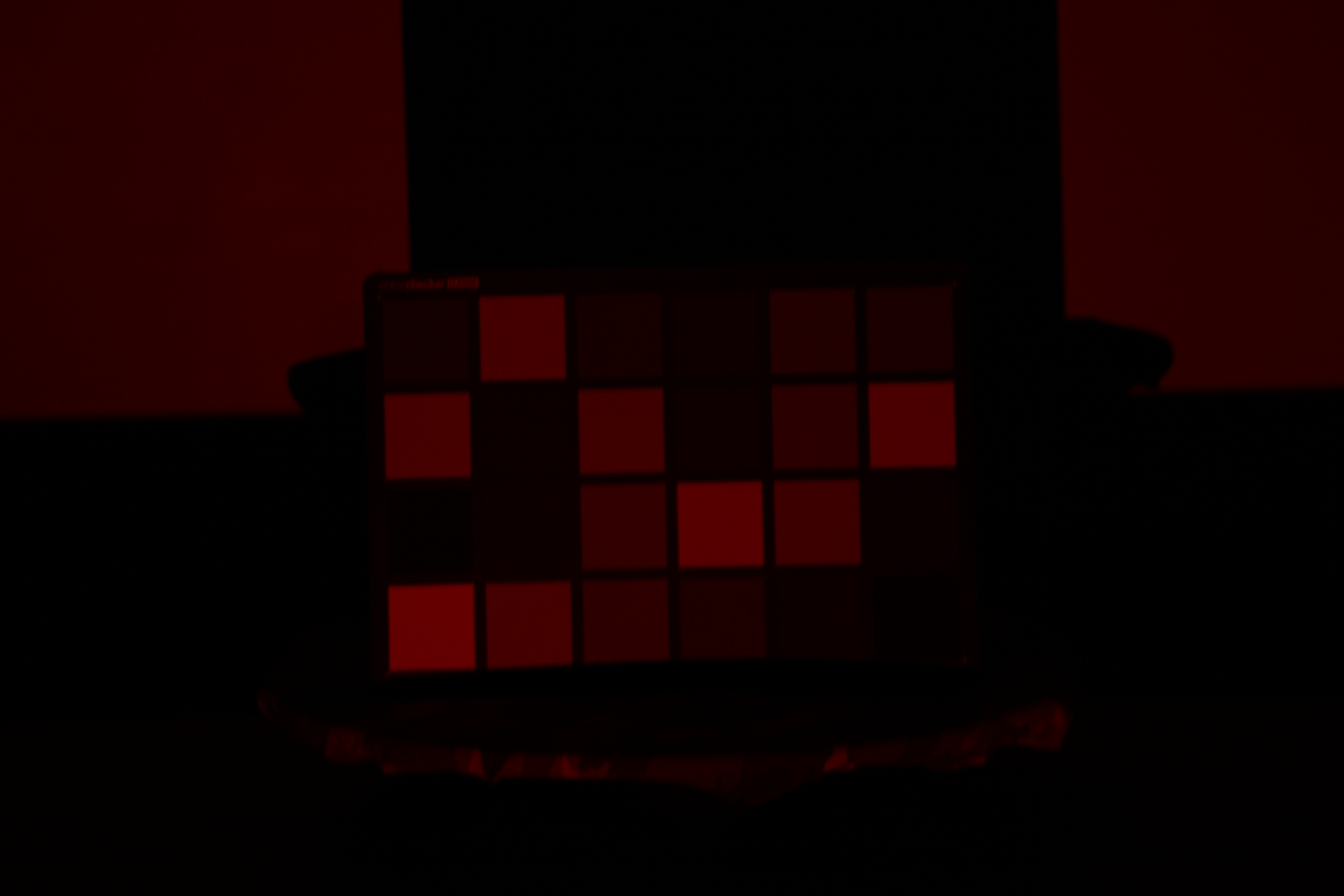}\\
        \textrm{\small{Red}}
    \end{minipage}
    \begin{minipage}[t]{0.133\linewidth}
        \centering
        \includegraphics[width =\linewidth]{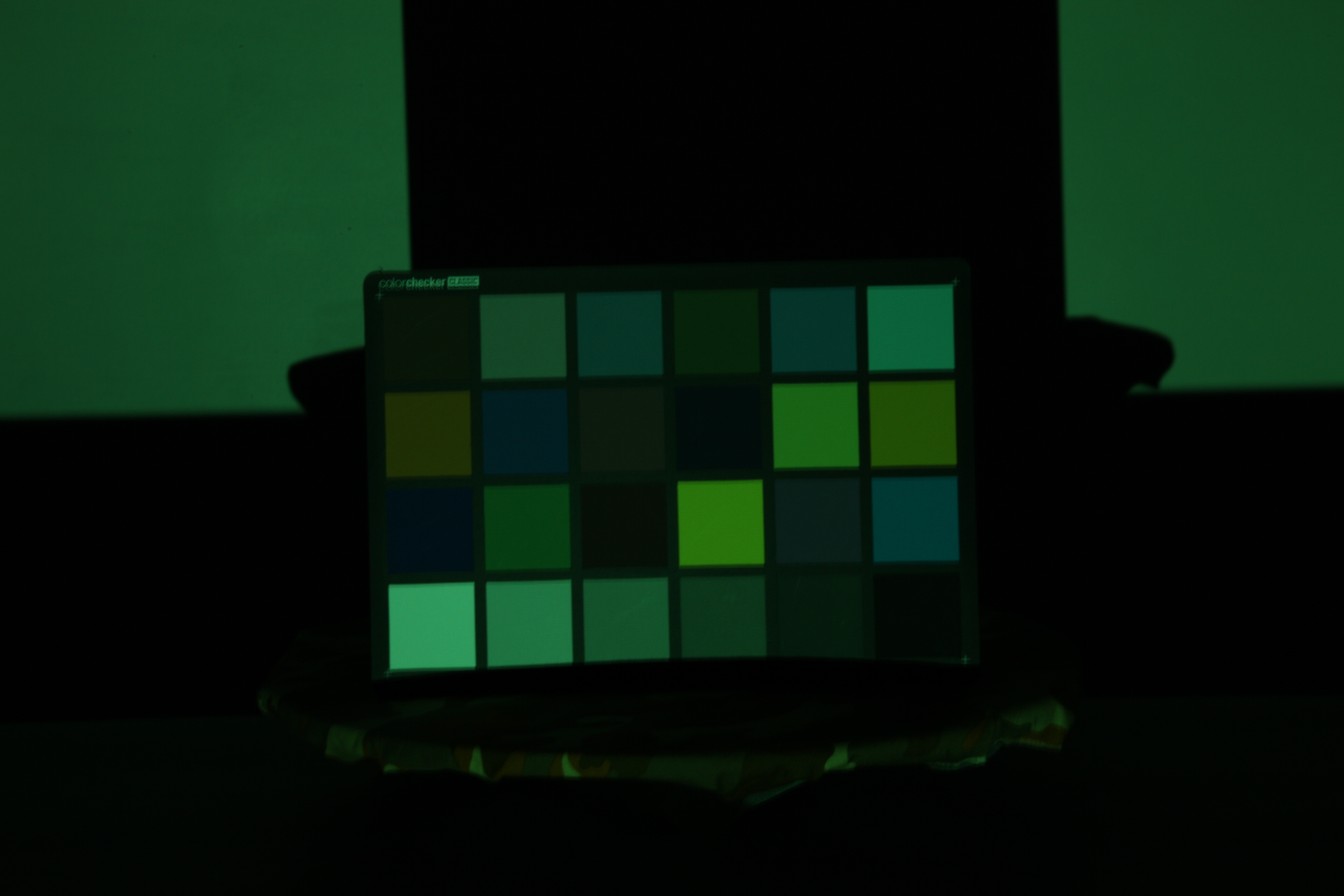}\\
        \textrm{\small{Green}}
    \end{minipage}
    \begin{minipage}[t]{0.133\linewidth}
        \centering
        \includegraphics[width =\linewidth]{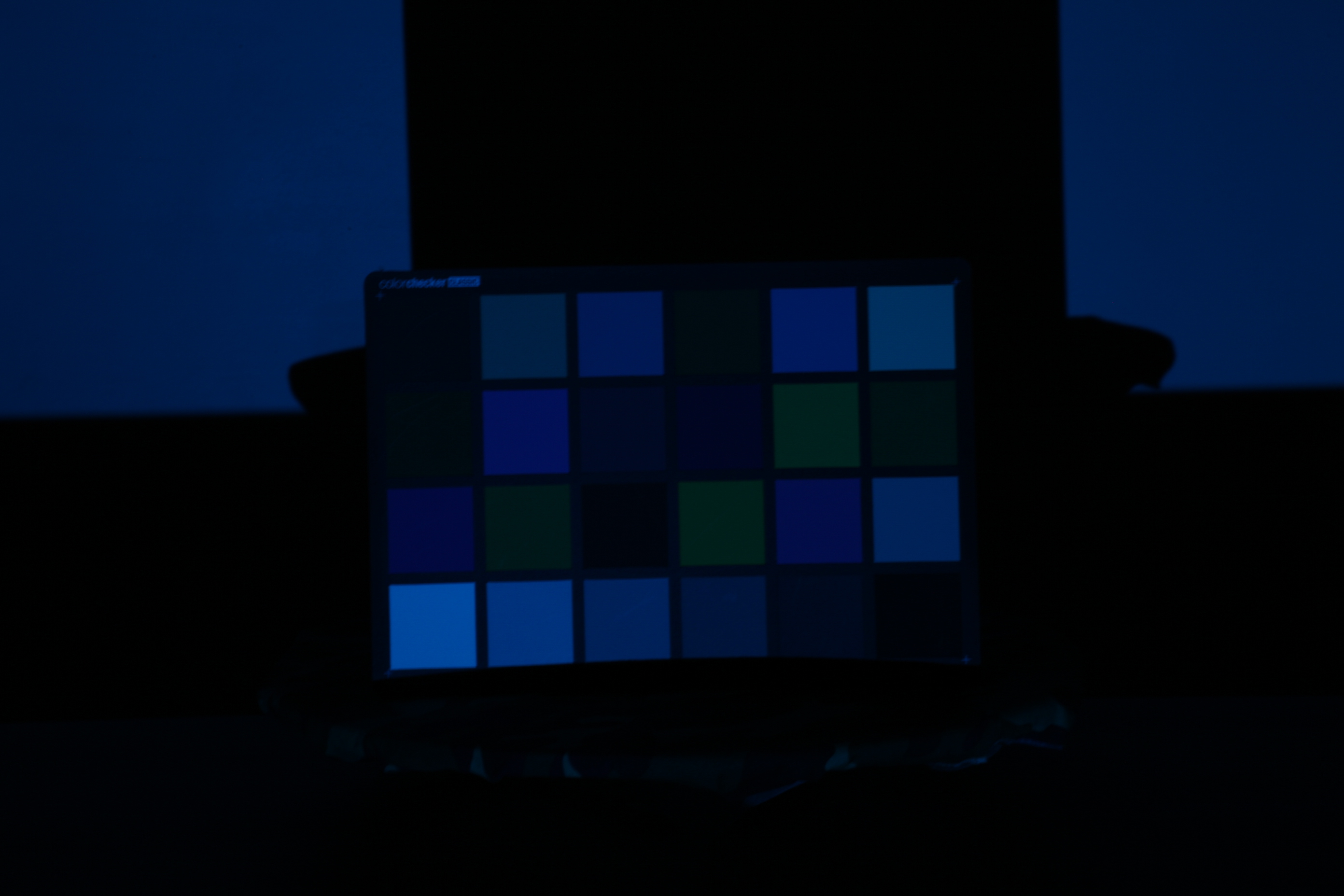}\\
        \textrm{\small{Blue}}
    \end{minipage}
    \begin{minipage}[t]{0.133\linewidth}
        \centering
        \includegraphics[width =\linewidth]{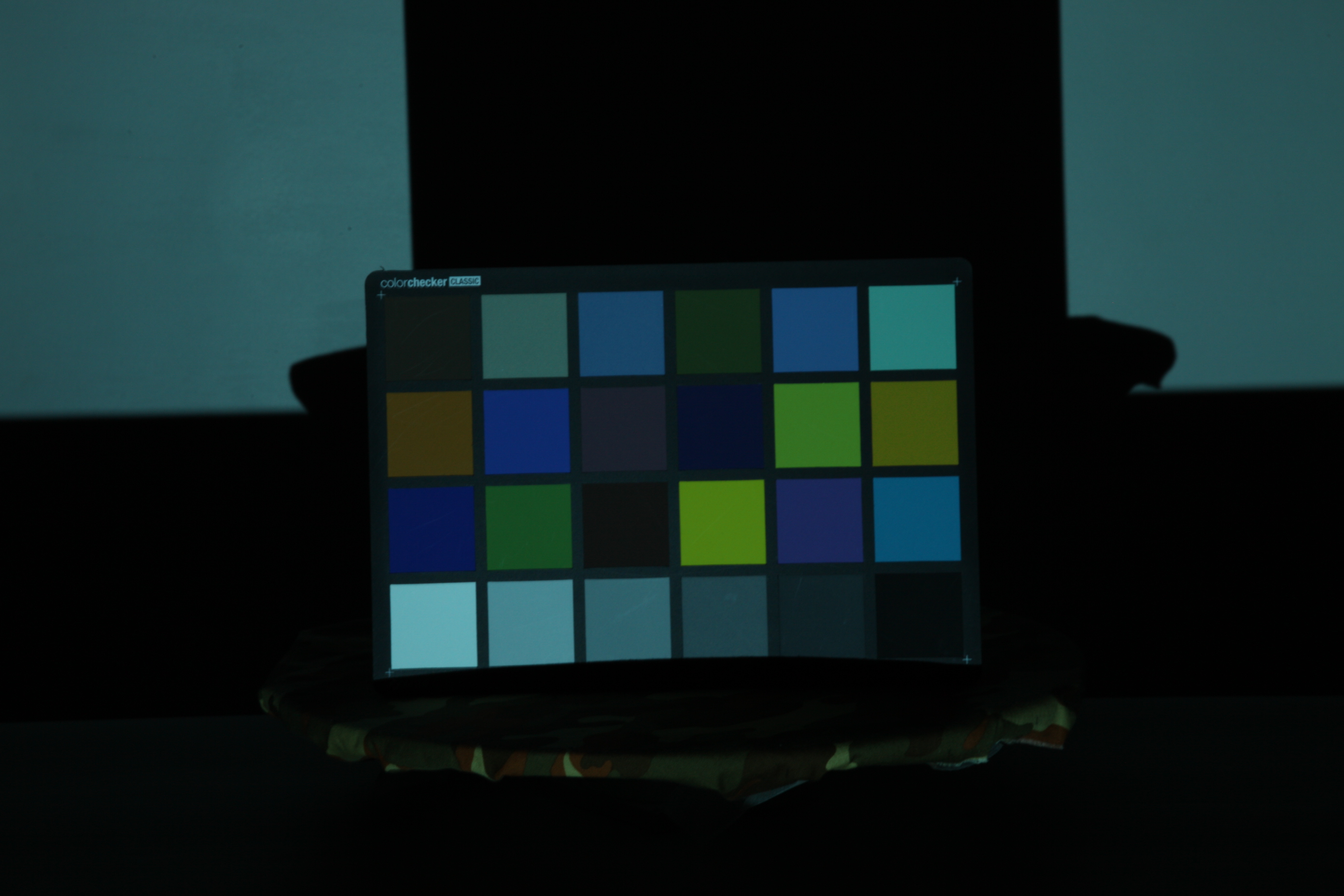}\\
        \textrm{\small{Cyan}}
    \end{minipage}
    \begin{minipage}[t]{0.133\linewidth}
        \centering
        \includegraphics[width =\linewidth]{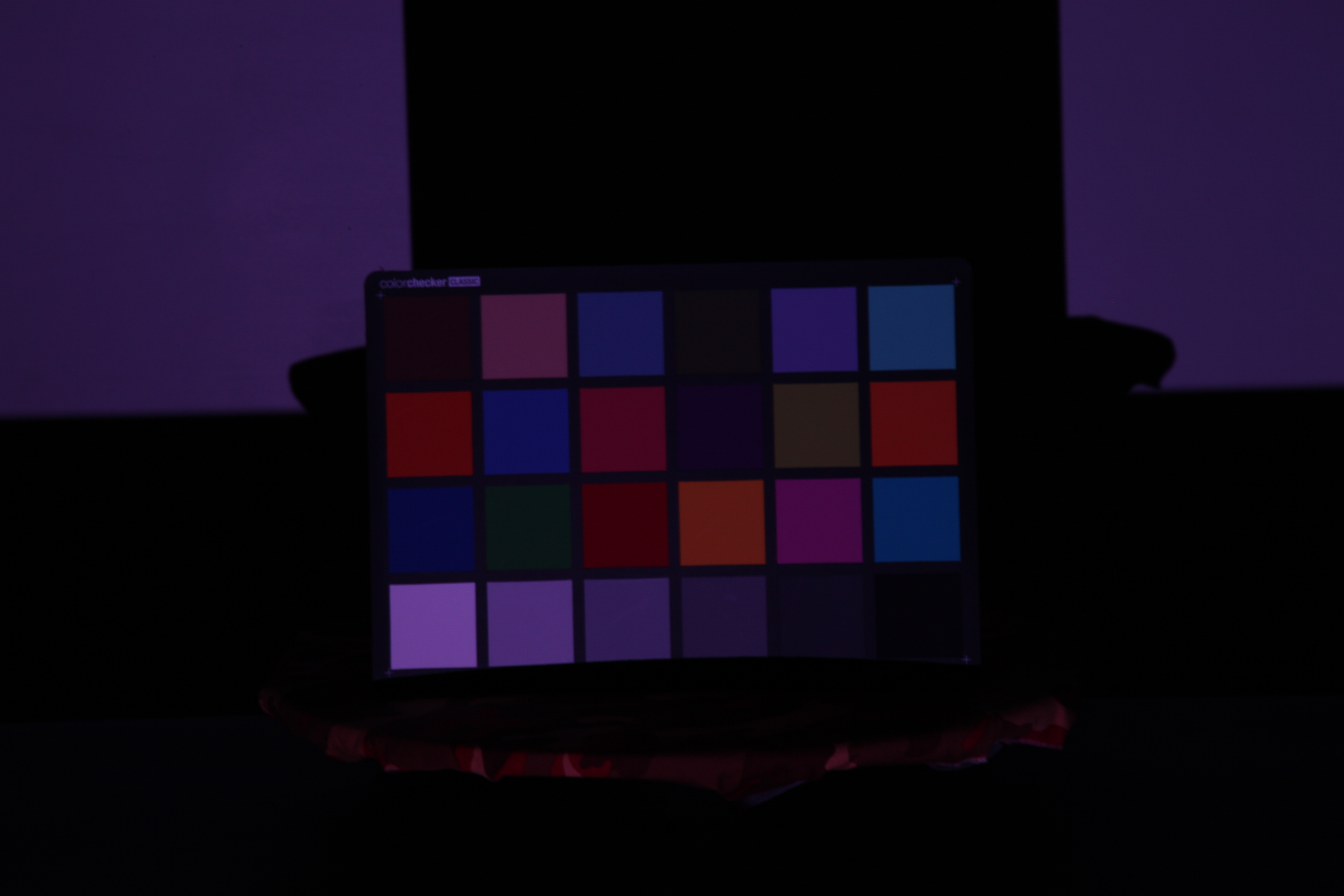}\\
        \textrm{\small{Magenta}}
    \end{minipage}
    \begin{minipage}[t]{0.133\linewidth}
        \centering
        \includegraphics[width =\linewidth]{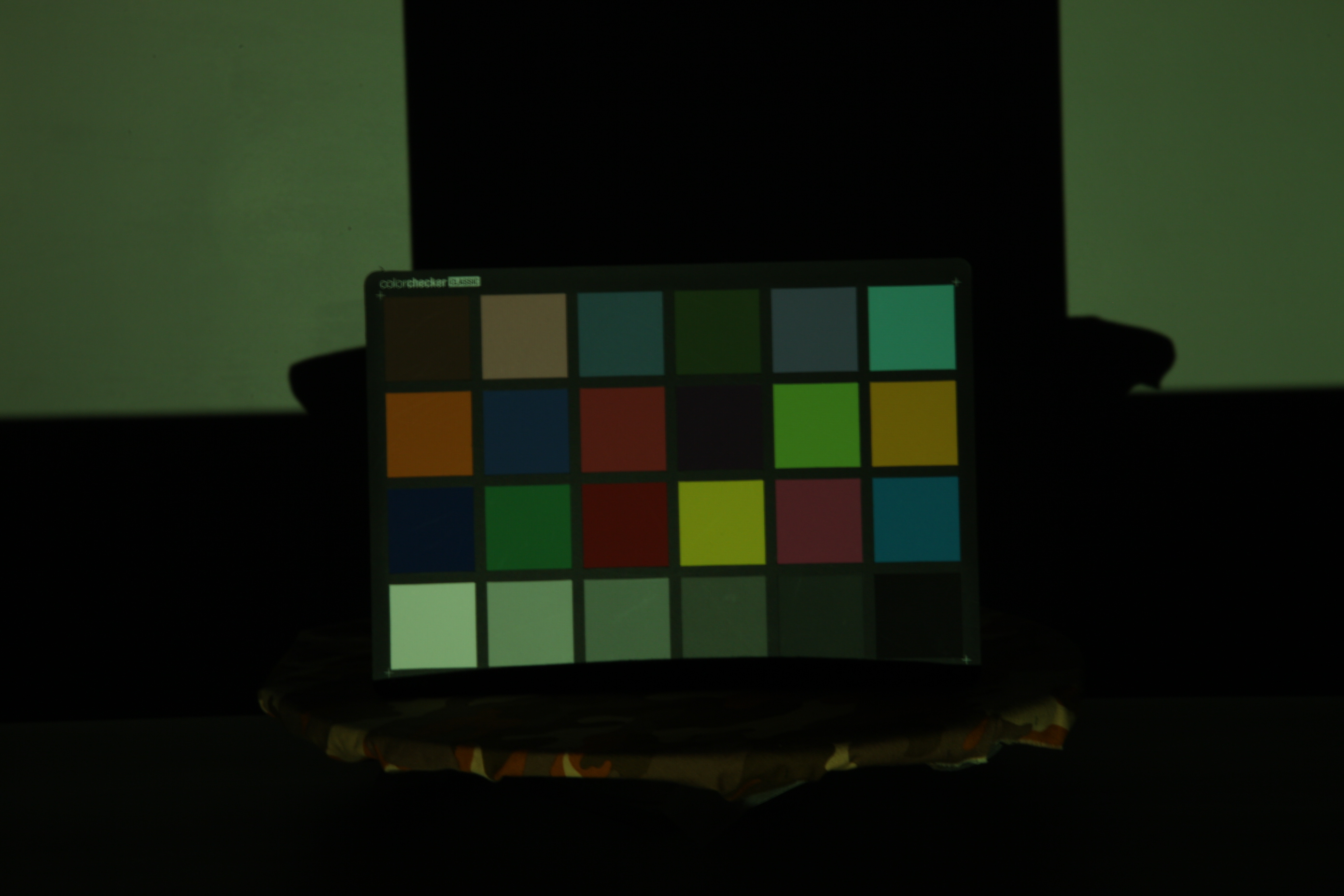}\\
        \textrm{\small{Yellow}}
    \end{minipage}
    \begin{minipage}[t]{0.133\linewidth}
        \centering
        \includegraphics[width =\linewidth]{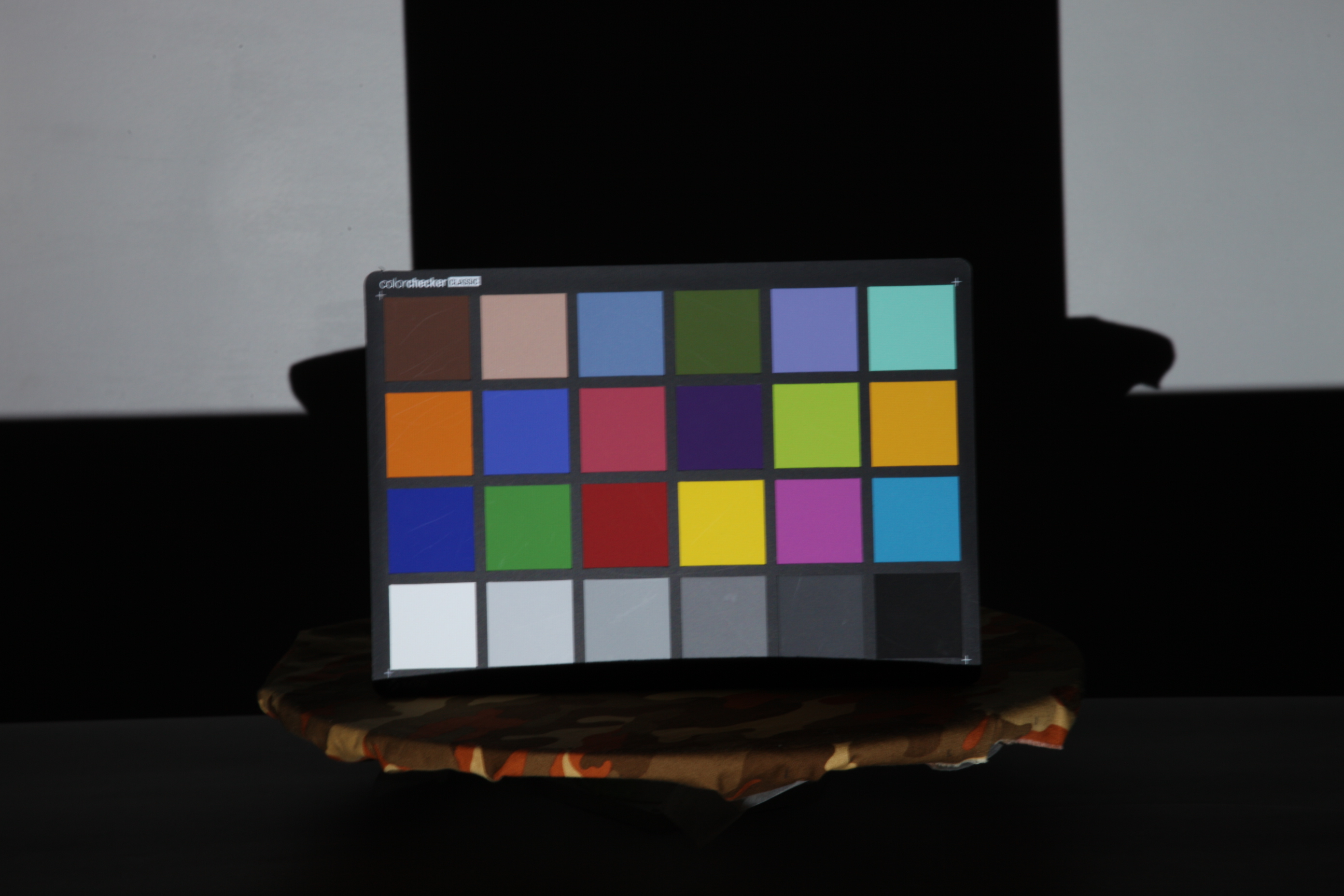}\\
        \textrm{\small{White}}
    \end{minipage}
    \end{tabular}
    \vspace{-1mm}
    \subcaption{Captured images under seven projected color illuminations}\label{fig:inputimages}
\end{minipage}\\ \vspace{10mm}

%%illumination result
\centering
\begin{minipage}{\hsize}
    \includegraphics[width=0.98\linewidth]{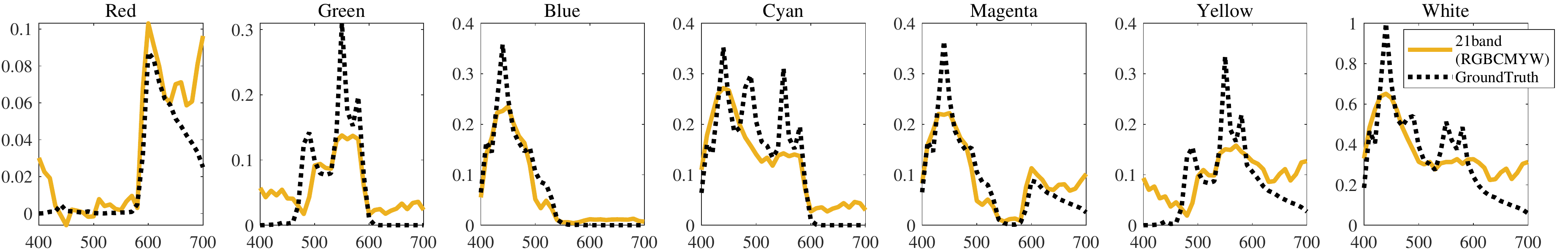}
    \subcaption{Estimation results of projector's SPDs}
    \label{fig:Ill_recover}
\end{minipage} \\ \vspace{10mm}

%%reflectance result
\centering
\begin{minipage}{\hsize}
    \includegraphics[width=0.98\linewidth]{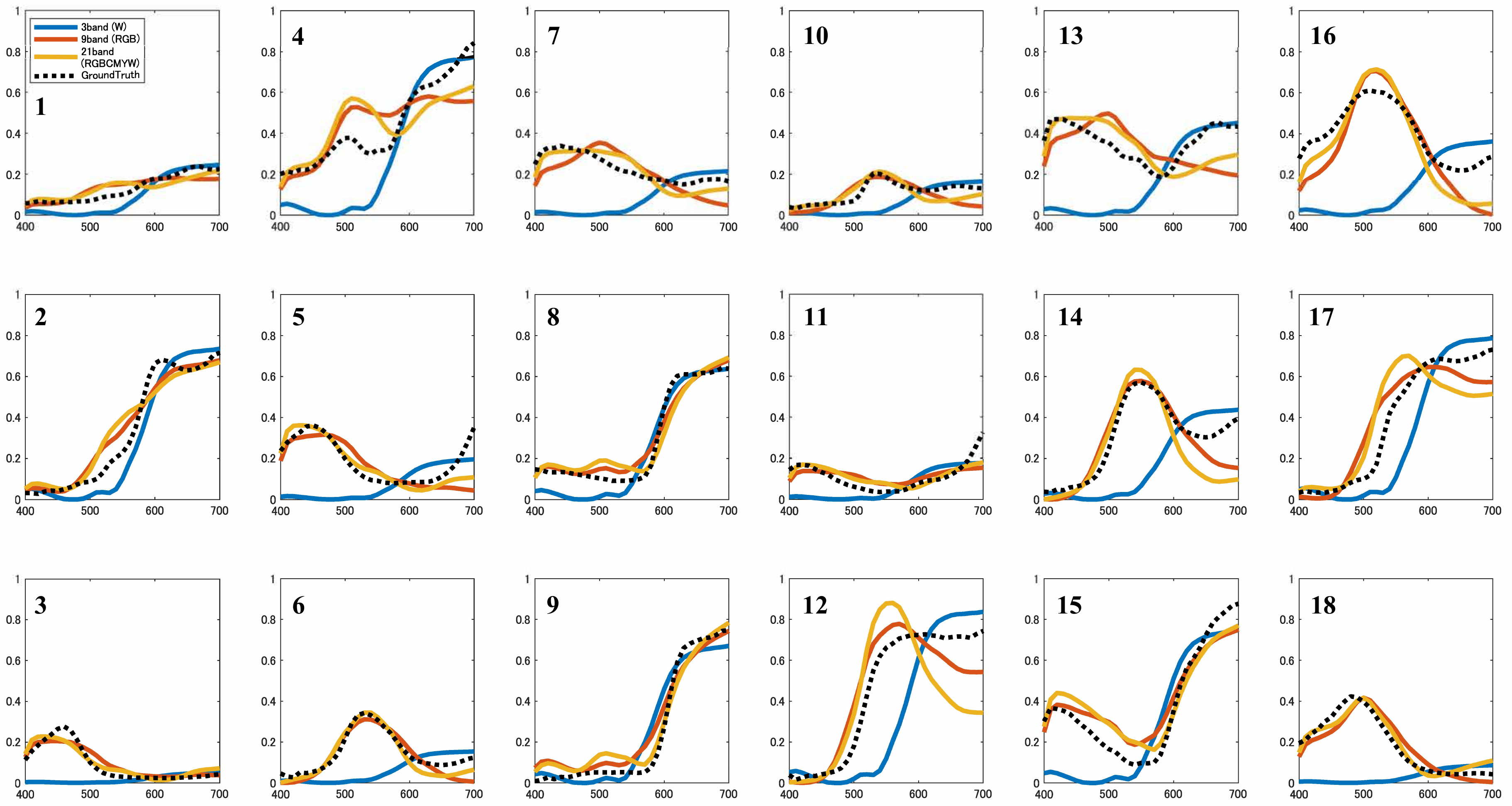}
    \subcaption{Spectral reflectance estimation results for 18 chromatic patches of the colorchart}
    \label{fig:Ref_revover}
\end{minipage}
\vspace{8mm}
\caption{Joint estimation results for X-Rite colorchart}
\vspace{4mm}
\label{fig:CCDC}
\end{figure*}

\subsection{Colorchart results}\label{sec:synthetic}
\vspace{1mm}
We evaluated the performance of the joint estimation using X-Rite Colorchart Classic images. Fig.~\ref{fig:CCDC}(a) shows the captured images under the seven projected illuminations. From the captured images, we sampled pixels from 18 chromatic patches and used those pixels for the joint estimation of the spectral reflectance and the projector's SPDs. Fig.~\ref{fig:RMSE_variation} shows the convergence analysis of the minimized cost. We can see that the cost converges after roughly 200 iterations of the spectral reflectance estimation and the projector's SPD estimation steps. 

Fig.~\ref{fig:CCDC}(c) shows the spectral reflectance estimation results for the 18 chromatic patches of the colorchart. We compared the joint estimation results of three cases using a different number of projected illuminations: 3-band imaging using only the white illumination, 9-band imaging using the red, the green, and the blue illuminations, and 21-band imaging using all the seven illuminations. As we can see from Fig.~\ref{fig:CCDC}(c), the spectral reflectance results using only the white illumination is very worse. This is because that the estimation of the projector's SPD from only the 3-band data is very difficult. On the other hand, the overall shapes of the spectral reflectances can be estimated in the 9-band and the 21-band cases. The average RMSE errors for these two cases are shown in Table~\ref{tab:RMSE_ref}, where the two cases show similar average RMSE errors. This is because that the cyan, the magenta, the yellow, and the white illuminations can be represented by the sum of three primary illuminations and thus the intrinsic dimension of the 21-band case is 9. However, the 21-band case could have the advantage of improving the robustness to the measurement noise by using more observations.

% ////////////////////////////////////////////
\begin{table*}[t!]
 \begin{center}
  \caption{RMSE result of spectral reflectance for each patch of the colorchart}
  \vspace{1mm}
  \begin{tabular}{|l||c|c|c|c|c|c|c|c|c|c|} \hline
        Patch & 1 & 2 & 3 & 4 & 5 & 6 & 7 & 8 & 9 & 10\\ \hline \hline
        3band & 0.046 & 0.074 & 0.122 & 0.185 & 0.186 & 0.149 & 0.203 & 0.076 & 0.087 & 0.077\\ \hline
        9band & 0.036 & $\bm{0.064}$ & 0.033 & 0.133 & 0.089 & 0.046 & 0.081 & $\bm{0.051}$ & 0.069 & 0.047\\ \hline
        21band &  $\bm{0.035}$ & 0.083 & $\bm{0.031}$ & $\bm{0.131}$ & $\bm{0.066}$ & $\bm{0.036}$ & $\bm{0.045}$ & 0.063 & $\bm{0.058}$ & $\bm{0.037}$ \\ \hline
  \end{tabular}\vspace{2mm}
  \begin{tabular}{|l||c|c|c|c|c|c|c|c||c|} \hline
        Patch & 11 & 12 & 13 & 14 & 15 & 16 & 17 & 18 & Average \\ \hline \hline
        3band & 0.083 & 0.225 & 0.261 & 0.232 & 0.181 & 0.366 & 0.142 & 0.227 & 0.1620 \\ \hline
        9band & 0.050 & $\bm{0.115}$ & 0.135 & $\bm{0.089}$ & 0.091 & 0.131 & $\bm{0.105}$ & 0.052 & 0.0787 \\ \hline
        21band & $\bm{0.041}$ & 0.210 & $\bm{0.102}$ & 0.132 & $\bm{0.086}$ & $\bm{0.116}$ & 0.143 & $\bm{0.036}$ & $\bm{0.0785}$  \\ \hline
  \end{tabular}
  \label{tab:RMSE_ref}
 \end{center}
\end{table*}

% ////////////////////////////////////////////
\begin{figure*}[t!]
\begin{center}
    \begin{tabular}{c}
    \begin{minipage}[t]{0.20\linewidth}
        \centering
        \includegraphics[width =\linewidth]{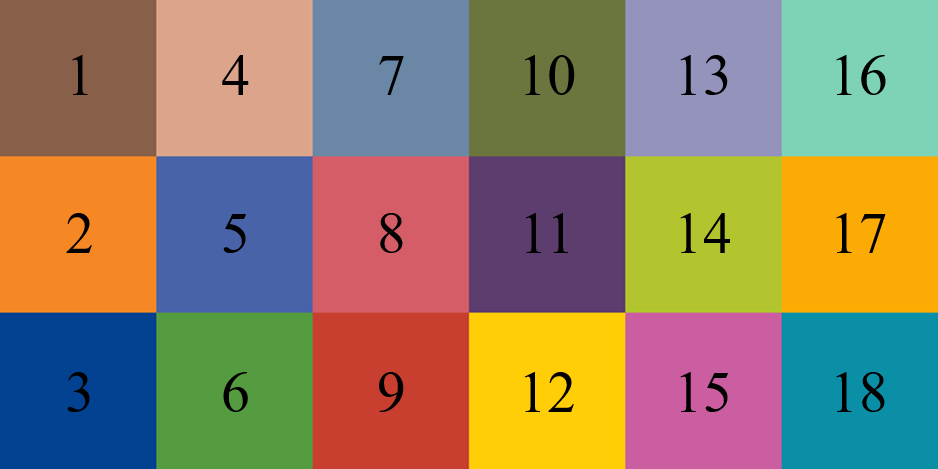}\\
        \textrm{\small{Patch numbering}}
    \end{minipage}
    % \hspace{3mm}
    \begin{minipage}[t]{0.20\linewidth}
        \centering
        \includegraphics[width =\linewidth]{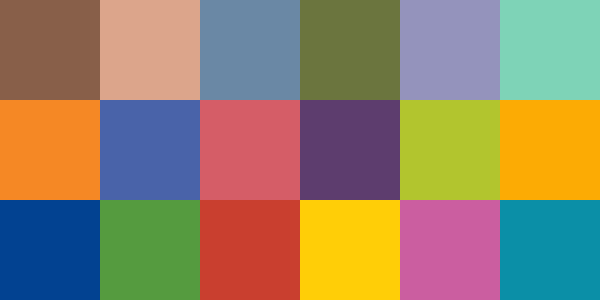}\\
        \textrm{\small{Ground truth}}
    \end{minipage}
    % \hspace{3mm}
    \begin{minipage}[t]{0.20\linewidth}
        \centering
        \includegraphics[width =\linewidth]{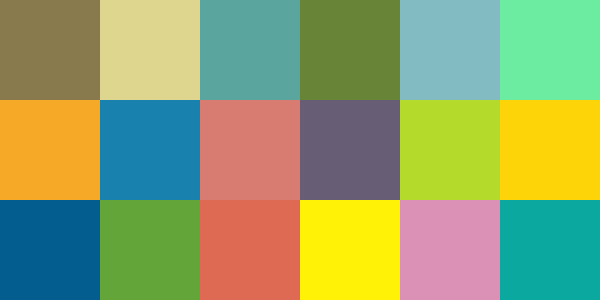}\\
        \textrm{\small{9band}}
    \end{minipage}
    % \hspace{3mm}
    \begin{minipage}[t]{0.20\linewidth}
        \centering
        \includegraphics[width =\linewidth]{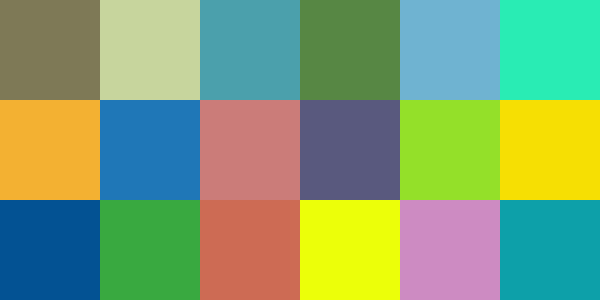}\\
        \textrm{\small{21band}}
    \end{minipage}
    % \hspace{3mm}
    \end{tabular}
\vspace{2mm}
\caption{Rendered sRGB images under the D65 illumination}
\vspace{5mm}
\label{fig:relighting}
\end{center}
\end{figure*}

% ////////////////////////////////////////////
\begin{table*}[t!]
 \begin{center}
  \vspace{5mm}
  \caption{CIE DE2000 color difference errors for each patch of the colorchart}
  \vspace{1mm}
  \begin{tabular}{|l||c|c|c|c|c|c|c|c|c|c|} \hline
        Patch & 1 & 2 & 3 & 4 & 5 & 6 & 7 & 8 & 9 & 10\\ \hline \hline
        9band & $\bm{24.5}$ & $\bm{15.61}$ & 18.36 & $\bm{33.92}$ & 21.96 & $\bm{2.84}$ & 26.99 & 10.25 & $\bm{9.45}$ & $\bm{6.64}$\\ \hline
        21band & 26.12 & 19.46 & $\bm{10.12}$ & 40.61 & $\bm{15.26}$ & 4.43 & $\bm{18.75}$ & $\bm{9.24}$ & 9.63 & 11.58 \\ \hline
  \end{tabular}\vspace{2mm}
  \begin{tabular}{|l||c|c|c|c|c|c|c|c||c|} \hline
        Patch & 11 & 12 & 13 & 14 & 15 & 16 & 17 & 18 & Average \\ \hline \hline
        9band & $\bm{16.60}$ & $\bm{12.12}$ & 29.64 & $\bm{4.19}$ & 12.29 & 13.12 & $\bm{17.64}$ & 17.51 & $\bm{16.29}$ \\ \hline
        21band & 16.75 & 19.67 & $\bm{28.82}$ & 11.02 & $\bm{11.20}$ & $\bm{10.11}$ & 22.88 & $\bm{8.99}$ & 16.37  \\ \hline
  \end{tabular}
  \label{tab:RMSE_ref_sRGB}
 \end{center}
\end{table*}

Fig.~\ref{fig:CCDC}(b) shows the projector's SPD estimation results when using all the seven illuminations. As we can see from the results, the overall shape of the SPDs can be estimated. However, the narrow-band peaks of the projector's SPDs cannot be accurately estimated. This is the limitation of the current projector's SPD basis functions that were obtained from a relatively small number of real projectors. We can also see that the estimation results for the longer part of the wavelengths (e.g., [650nm, 700nm]) are worse than the results for the other wavelengths. This trend can be seen also in the spectral reflectance estimation results of Fig.~\ref{fig:CCDC}(c). This can be due to the low spectral power of the projector and the low camera sensitivity at those wavelengths. Even though the accuracy of our joint estimation is still not very accurate, it could be further improved by deriving better basis functions for the projector's SPDs.

Fig.~\ref{fig:relighting} shows rendered sRGB images using ground-truth and estimated spectral reflectances. Table~\ref{tab:RMSE_ref_sRGB} shows the CIE DE2000 color difference errors for each patch. In terms of color difference, the 9-band case slightly provides better color accuracy on average.

%
% ======= Contributions ==================================================
%
\section{Conclusion}
\vspace{1mm}
In this paper, we have introduced a method for spectral reflectance estimation using a standard RGB camera and an off-the-shelf projector with unknown SPD. Based on a newly collected projector's SPD database, we have modeled the projector's SPD by a low-dimensional basis model. Then, we have solved the cost minimization problem to jointly estimate the spectral reflectances and the projector's SPDs. Experimental results have demonstrated that the overall shapes of the projector's SPDs as well as the spectral reflectances can be estimated using more than 9-band data.
To further improve the performance of the joint estimation, in future work, we will try to collect more projector's SPD data to derive better basis functions. Applying the joint estimation to the spectral 3D acquisition system of~\cite{li2019pro} would be one of interesting future directions.

\section{Acknowledgments} 
This work was partly supported by JSPS KAKENHI Grant Number 17H00744.

%%%%%%%%%%%%%%%%%%%%%%%%%%%%%%%%%%
% Reference Preparation
%%%%%%%%%%%%%%%%%%%%%%%%%%%%%%%%%%
% \newpage
\bibliographystyle{IEEEtran}
\bibliography{refs}

% Generated by IEEEtran.bst, version: 1.12 (2007/01/11)
\begin{thebibliography}{10}
\providecommand{\url}[1]{#1}
\csname url@samestyle\endcsname
\providecommand{\newblock}{\relax}
\providecommand{\bibinfo}[2]{#2}
\providecommand{\BIBentrySTDinterwordspacing}{\spaceskip=0pt\relax}
\providecommand{\BIBentryALTinterwordstretchfactor}{4}
\providecommand{\BIBentryALTinterwordspacing}{\spaceskip=\fontdimen2\font plus
\BIBentryALTinterwordstretchfactor\fontdimen3\font minus
  \fontdimen4\font\relax}
\providecommand{\BIBforeignlanguage}[2]{{%
\expandafter\ifx\csname l@#1\endcsname\relax
\typeout{** WARNING: IEEEtran.bst: No hyphenation pattern has been}%
\typeout{** loaded for the language `#1'. Using the pattern for}%
\typeout{** the default language instead.}%
\else
\language=\csname l@#1\endcsname
\fi
#2}}
\providecommand{\BIBdecl}{\relax}
\BIBdecl

\bibitem{liang2012advances}
H.~Liang, ``Advances in multispectral and hyperspectral imaging for archaeology
  and art conservation,'' \emph{Applied Physics A}, vol. 106, no.~2, pp.
  309--323, 2012.

\bibitem{qin2013hyperspectral}
J.~Qin, K.~Chao, M.~S. Kim, R.~Lu, and T.~F. Burks, ``Hyperspectral and
  multispectral imaging for evaluating food safety and quality,'' \emph{Journal
  of Food Engineering}, vol. 118, no.~2, pp. 157--171, 2013.

\bibitem{gat2000imaging}
N.~Gat, ``Imaging spectroscopy using tunable filters: {A} review,'' \emph{Proc.
  of SPIE}, vol. 4056, pp. 50--64, 2000.

\bibitem{Monno}
Y.~Monno, S.~Kikuchi, M.~Tanaka, and M.~Okutomi, ``A practical one-shot
  multispectral imaging system using a single image sensor,'' \emph{IEEE Trans.
  on Image Processing}, vol.~24, no.~10, pp. 3048--3059, 2015.

\bibitem{Park}
J.~Park, M.~Lee, M.~D. Grossberg, and S.~K. Nayar, ``Multispectral imaging
  using multiplexed illumination,'' \emph{Proc. of IEEE Int. Conf. on Computer
  Vision (ICCV)}, pp. 1--8, 2007.

\bibitem{Cui}
C.~Cui, H.~Yoo, and M.~Ben-Ezra, ``Multi-spectral imaging by optimized wide
  band illumination,'' \emph{Int. Journal of Computer Vision}, vol.~86, pp.
  140--151, 2010.

\bibitem{Han2}
S.~Han, I.~Sato, T.~Okabe, and Y.~Sato, ``Fast spectral reflectance recovery
  using {DLP} projector,'' \emph{Int. Journal of Computer Vision}, vol. 110,
  no.~2, pp. 172--184, 2014.

\bibitem{li2019pro}
C.~Li, Y.~Monno, H.~Hidaka, and M.~Okutomi, ``{Pro-Cam SSfM}:
  {P}rojector-camera system for structure and spectral reflectance from
  motion,'' \emph{Proc. of IEEE Int. Conf. on Computer Vision (ICCV)}, pp.
  2414--2423, 2019.

\bibitem{Parkkinen}
J.~Parkkinen, J.~Hallikainen, and T.~Jaaskelainen, ``Characteristic spectra of
  munsell colors,'' \emph{Journal of the Optical Society of America A}, vol.~6,
  no.~2, pp. 318--322, 1989.

\bibitem{Maloney}
L.~T. Maloney, ``Evaluation of linear models of surface spectral reflectance
  with small numbers of parameters,'' \emph{Journal of the Optical Society of
  America A}, vol.~3, no.~10, pp. 1673--1683, 1986.

\bibitem{Monno2}
Y.~Monno, M.~Tanaka, and M.~Okutomi, ``Direct spatio-spectral datacube
  reconstruction from raw data using a spatially adaptive spatio-spectral
  basis,'' \emph{Proc. of SPIE}, vol. 8660, pp. 866\,003--1--866\,003--8, 2013.

\bibitem{Blasinski}
H.~Blasinski and J.~Farrell, ``Computational multispectral flash,'' \emph{Proc.
  of IEEE Int. Conf. on Computational Photography (ICCP)}, pp. 1--10, 2017.

\bibitem{Oh}
S.~W. Oh, M.~S. Brown, M.~Pollefeys, and S.~J. Kim, ``Do it yourself
  hyperspectral imaging with everyday digital cameras,'' \emph{Proc. of IEEE
  Conf. on Computer Vision and Pattern Recognition (CVPR)}, pp. 2461--2469,
  2016.

\bibitem{Jiang}
J.~Jiang, D.~Liu, J.~Gu, and S.~S{\"u}sstrunk, ``What is the space of spectral
  sensitivity functions for digital color cameras?'' \emph{Proc. of IEEE Winter
  Conf. on Applications of Computer Vision (WACV)}, pp. 168--179, 2013.

\end{thebibliography}

\end{document}